\documentclass[aps,prl,twocolumn,showpacs,superscriptaddress,groupedaddress]{revtex4-1}  
\usepackage{graphicx}  
\usepackage{dcolumn}   
\usepackage{bm}        
\usepackage{amssymb}   
\usepackage{framed}
\usepackage{amsmath}

\begin{document}

\widetext
\title{Semi-local Exchange Energy Functional For Two-Dimensional Quantum Systems: A Step Beyond
Generalized Gradient Approximations}

\author{Subrata Jana}
\affiliation{School of Physical Sciences, National Institute of Science Education and Research,
Bhubaneswar 752050, Homi Bhava National Institute, INDIA}
\author{Prasanjit Samal}
\affiliation{School of Physical Sciences, National Institute of Science Education and Research,
Bhubaneswar 752050, Homi Bhava National Institute, INDIA}
                             
\date{\today}

\begin{abstract}
Semi-local density functionals for the exchange-correlation energy of electrons are extensively
used as it produce realistic and accurate results for finite and extended systems. The choice of
techniques play crucial role in constructing such functionals of improved accuracy and efficiency.
An accurate and efficient semi-local exchange energy functional in two dimensions is constructed
by making use of the corresponding hole based on the density matrix expansion. The exchange hole
involved is localized under the generalized coordinate transformation and satisfies all the relevant
constraints. Comprehensive testing and excellent performance of the functional is demonstrated versus
exact exchange results. The functional also achieves remarkable accuracy by substantially reducing
the errors present in the local and non-empirical density functionals proposed so far for two
dimensional systems. The underlying principles involved in the functional construction are physically
appealing and practically useful for developing range separated and non-local functionals in two
dimensions.
\end{abstract}

\maketitle

Density-functional theory(DFT) \cite{hk64,ks65} is most successful in addressing the complex effects 
due to electron-electron interactions. Tremendous advances beyond the local density approximation(LDA) 
have been achieved through the development of accurate non-local, semi-local and hybrid exchange-
correlation(XC) functionals \cite{b83,jp85,pw86,b88,br89,b3pw91,pbe96,kos,vsxc98,hcth,tsuneda,tpss,
mO6l,revtpss,tbmbj,scan15,tm16}. However, cutting edge research in low dimensions have also gained 
momentum as far as the theoretical and experimental findings \cite{kat,rm} are concerned. In spite 
of the promising applications in three dimensions(3D), the dimensional crossover of the XC energy 
functional from a 3D to two-dimensional(2D) regime, has still remained one of the most difficult open 
problems \cite{klnlhm,ccfs}. Albeit wide use of DFT in 2D still demands potentiality in this direction. 
So the systematic DFT calculations and proper explanations of numerous properties of low-dimensional 
systems that range from atomistic to artificial structures e.g., quantum dots, modulated semiconductor 
layers and surfaces, quantum Hall systems, spintronic devices, quantum rings, and artificial graphene 
poses great challenge. Thus, the construction of accurate non-local and semilocal XC functionals to 
appropriately describe systems in 2D is an enthralling and growing research field. In this regard, the 
first step among the available methods is the well-known 2D-LDA \cite{rk}. The 2D-LDA combined with 
the 2D correlation \cite{tc,amgb}, lead to intriguing results and establishes its superiority over 
quantum Monte Carlo simulations \cite{hser} result. In recent years, advances have been made beyond 
2D-LDA e.g., generalized gradient approximations(2D-GGAs) \cite{prhg,prvm,prg,pr1,prp,rp,sr,pr2,rpvm,
prlvm,vrmp} which perform in a more excellent manner. Not only that, several correlation functionals 
compatible with the 2D-GGAs are also constructed \cite{rp,sr,pr2, rpvm,prm,prpg,rpp}.  

In principle, the exchange functionals can be constructed from the exchange hole. In 3D, it's done 
by making use of Taylor series expansion \cite{b83,b3pw91}, real space cutoff procedure \cite{jp85}, 
modeling the exchange hole \cite{br89} and the density matrix expansion(DME) based on general coordinate 
transformation \cite{kos,vsxc98,hcth,tsuneda,mO6l,tm16}. It is to note that the Taylor series expansion 
method has been applied to construct 2D-GGA \cite{prvm}. However, unlike Taylor expansion, DME \cite{nv1,
kos,vsxc98,mO6l,tm16} based approaches are not only correct for small separation limit, but do converge 
in the large separation limit \cite{tm16} and recover the correct uniform gas behavior. Prompted by these, 
we have formulated the 2D counterpart of the above DME based exchange energy functional. Advance DME 
techniques will be proposed for constructing the exchange hole and the corresponding energy functional. 
Then, the functional will be bench-marked against the optimized effective potential(OEP) based exact-
exchange(EXX) \cite{kli}, local and gradient approximations for 2D systems \cite{prvm,vrmp,prhg}. The 
OEP based EXX functional is used as reference because it's the most accurate approach which is routinely 
applied for studying quantum dots \cite{hkprg}. Further, the newly constructed functional will be applied 
to study few electron trapped inside parabolic and Gaussian quantum dots.
 
\begin{figure}
\begin{center}
\includegraphics[width=3.4in,height=2.0in,angle=0.0]{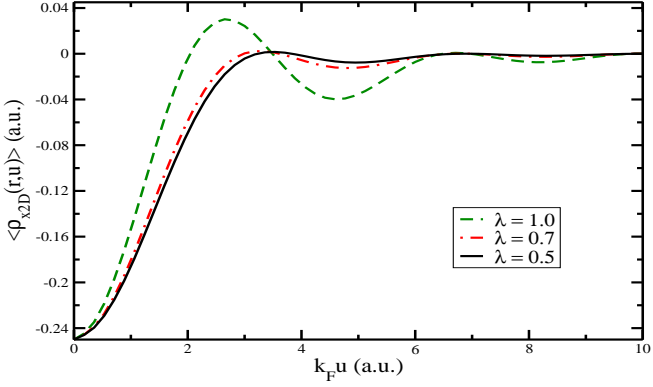} 
\caption{The Eq.(\ref{sec3eq6}) is plotted (with $j_p=0$) for two non-interacting electrons with density 
$\rho(r)=\frac{2}{\pi}\exp(-2r^2)$ parabolically confined in 2D. Shown are the exchange holes at the 
reference point $r = 0.5~a.u.$ for several $\lambda$ values, where $r$ be the radial distance from the 
origin. For $\lambda = 1.0$, some portion of the exchange hole is $+ve$ indicating that the exchange hole 
violates the negativity property and will underestimate the magnitude of exchange energy. This implies the 
requirement for normalization of exchange hole (see ref. \cite{tsp03} for details).}
\label{fig1}
\end{center}
\end{figure}

\begin{figure}
\begin{center}
\includegraphics[width=3.8 in,height=3.0in,angle=0.0]{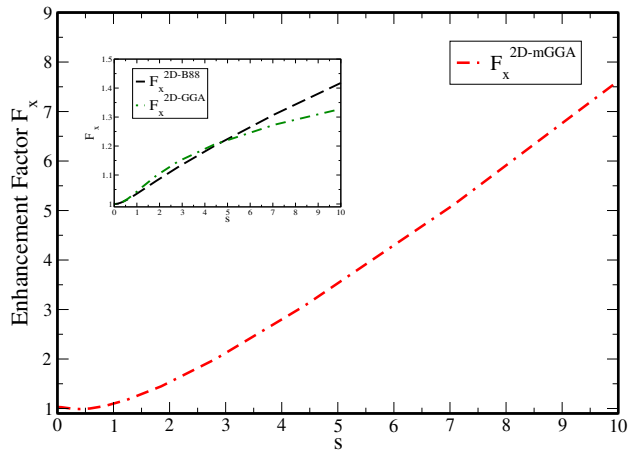} 
\end{center}
\caption{The enhancement factor $F_x^{2D-mGGA}$ (with $j_p =0$) is plotted as a functional of $s$ for $2$ 
electrons confined in a parabolic quantum dot with confinement strength $\omega = 1$. In the left corner, 
we have shown the enhancement factors of 2D-GGA \cite{prvm} and 2D-B88 \cite{vrmp} for comparison.}
\label{fig2}
\end{figure}

\begin{figure}
\begin{center}
\includegraphics[width=3.4in,height=2.0in,angle=0.0]{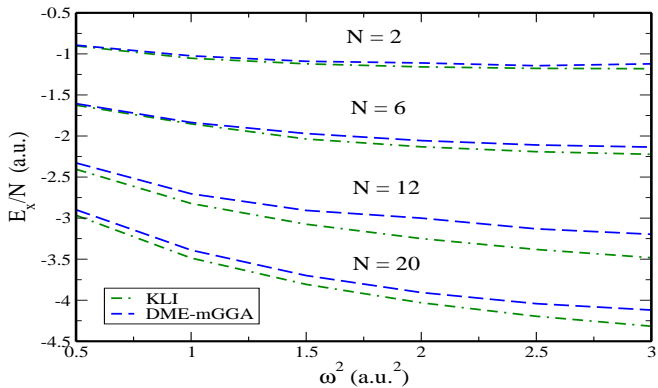} 
\end{center}
\caption{Shown in the figure, the exchange energy per electron (in a.u.) plotted versus $\omega^2$ for 
a series of Gaussian quantum dots with $N$ electrons and confinement strength $\omega$.}
\label{fig3}
\end{figure}

The exchange energy is nothing but the electrostatic interaction between the electron located at $\vec{r}$ 
and the exchange hole at $\vec{r}+\vec{u}$ surrounding it. Thus, the spin-unpolarized exchange functional 
in 2D is defined as
\begin{equation}
E^{2D}_x[\rho]=\frac{1}{2}\int~d^2r\int\frac{\rho(\vec{r})\rho_x(\vec{r},\vec{r}+\vec{u})}
{u}~d^2u,
\label{sec2eq1}
\end{equation}
where $\rho_x(\vec{r},\vec{r}+\vec{u})$ be the exchange hole surrounding the electron at $\vec r$ and is 
given by
\begin{equation}
\rho_x(\vec{r},\vec{r}+\vec{u})=-\frac{|\varGamma(\vec{r},\vec{r}+\vec{u})|^2}{2\rho(\vec{r})}
\label{sec2eq2}
\end{equation}
with the density matrix $\varGamma(\vec{r},\vec{r}+\vec{u})=2\sum_{i}^{occ}\psi_i^*(\vec{r})\psi_i(\vec{r}
+\vec{u})$ and the Kohn-Sham (KS) orbitals $\psi_i$ . The exchange hole obeys two important properties: 
(i) the normalization sum rule: $\int~\rho_x(\vec{r},\vec{r}+\vec{u})~d^2u = -1$ and (ii) the negativity 
constraint: $\rho_x(\vec{r},\vec{r}+\vec{u})\leq0$. Now, under general coordinate transformation (i.e. 
$(\vec{r}_1,\vec{r}_2)\to(\vec{r}^\lambda,u)$), where $\vec{r}^\lambda=\lambda\vec{r}_1+(1-\lambda)\vec{r}_2$, 
the above exchange functional reduces to
\begin{equation}
E^{2D}_x=\frac{1}{2}\int~d^2r^\lambda\rho(\vec{r}^\lambda)\int\frac{\rho^t_{x2D}(\vec{r}^\lambda,u)}
{u}~d^2u,
\label{sec2eq4}
\end{equation}
where $\rho^t_{x2D}$ is the transformed exchange hole defined by
\begin{equation}
\rho^t_{x2D}=-\frac{|\varGamma_{1t}^{2D}(\vec{r}^\lambda-(1-\lambda)\vec{u},\vec{r}^{\lambda}+\lambda
\vec{u})|^2}{2\rho(\vec{r})}
\label{sec2eq5}
\end{equation}
with $\varGamma_{1t}^{2D}$ be the KS single particle density matrix. The real parameter, $\lambda$ can take 
values $1/2 \to 1$ (or, $0 \to 1/2$). The conventional and on top exchange holes (which is maximally localized 
in 2D \cite{tsp03}) correspond to $\lambda = 1$ and $\lambda = \frac{1}{2}$ respectively. So the transformed 
single particle KS density matrix around $u=0$ becomes
\begin{equation}
\begin{split}
\varGamma_{1t}^{2D}(\vec{r},\vec{u})=e^{\vec{u}.[-(1-\lambda)\vec{\nabla}_1+\lambda\vec{\nabla}_2]}
\varGamma_{1t}^{2D}(\vec{r},\vec{u})|_{\vec{u}=0}\\
=e^{\vec{u}.[-(1-\lambda)\vec{\nabla}_1+\lambda\vec{\nabla}_2]}
\sum_i^{occ}\Psi^{*}_i(\vec{r}^\lambda
-(1-\lambda)\vec{u})\Psi_i(\vec{r}^{\lambda}+\lambda\vec{u})|_{\vec{u}=0}~,
\label{sec2eq6}
\end{split}
\end{equation}
where $\vec{\nabla}_1$ and $\vec{\nabla}_2$ operate on $\Psi^{*}_i$ and $\Psi_i$ respectively. The exchange 
energy, $E^{2D}_x$ involves cylindrical average of the exchange hole $\langle\rho_x(\vec{r},\vec{r}+\vec{u})
\rangle_{cyl}$ over the direction of $\vec{u}$ i.e.
\begin{equation}
\langle\rho_x(\vec{r},\vec{r}+\vec{u})\rangle_{cyl}=\int\rho_x(\vec{r},\vec{r}+\vec{u})~\frac{d\Omega_u}{2\pi}~.
\label{sec2eq3}
\end{equation}
On taking the cylindrical average of the density matrix given in Eq.(\ref{sec2eq6}) after it's Taylor series 
expansion yields the correct small $u$ behavior, i.e.
\begin{equation}
\begin{split}
\langle\rho^t_{x2D}\rangle = -\frac{\rho(\vec{r})}{2}-\frac{1}{4}\Big[\Big(\lambda^2-\lambda+\frac{1}{2}
\Big)\nabla^2\rho(\vec{r})-2\tau\\
+\frac{1}{4}\Big(2\lambda-1\Big)^2\frac{|\vec{\nabla}\rho(\vec{r})|^2}
{\rho(\vec{r})}\Big]u^2~.
\end{split}
\label{sec2eq7}
\end{equation} 
The expression in Eq.(\ref{sec2eq7}) was originally proposed for the conventional exchange hole in 3D 
\cite{b83} and then extended to 2D \cite{prvm}. But, it failed to recover the uniform density limit. 
In order to recover it, the whole term was multiplied by the exchange hole of uniform electron gas 
\cite{prvm}. Whereas, here in this work, all the above deficiencies are accounted through the proposed 
novel approach based on DME. As a matter of which, it quite rightly obtains: (i) the correct uniform 
density limit, (ii) the cylindrically averaged exchange hole similar to that given in Eq.(\ref{sec2eq7}) 
when terms up to $u^2$ will be considered and (iii) the large $u$-limit (i.e. $0$ to $\infty$ integral 
limit of $u$) that converges without considering any cutoff procedure. Now, to construct the desired 
semilocal functional, we begin by considering the DME in Eq.(\ref{sec2eq6}) along with the following 
plane wave expansion in terms of the {\it{Bessel}} and {\it{Hypergeometric}} functions. So
\begin{equation}
e^{\frac{kucos\phi y}{k}} = {\mathcal{A}} + {\mathcal{B}}~,
\label{sec3eq1}
\end{equation}
where
\begin{eqnarray}
{\mathcal{A}} &=& \frac{2}{ku}\sum_{n=0}^\infty(-1)^n(2n+1)J_{2n+1}(ku)C_{2n}^1\Big(-i\frac{ycos\phi}{k}\Big)
\nonumber\\
{\mathcal{B}} &=& \frac{2}{ku^2}\sum_{n=0}^\infty(-1)^n(2n+1)J_{2n+1}(ku)\frac{1}{2\cos\phi}\nonumber\\
&&~~~~~~~~~~~~~~~~~~~~~~~~~~~~~\times\frac{\partial }{\partial y}\Big[C_{2n}^1\Big(-i\frac{ycos\phi}{k}\Big)\Big]
\label{sec3eq2}
\end{eqnarray}
and $\phi$ be the azimuthal angle. The polynomial, $C_{2n}^{m}$ is expressed as
\begin{equation}
C^{m}_{2\nu}(x)=(-1)^{\nu}
\begin{pmatrix}
           \nu+m-1 \\
           \nu
\end{pmatrix}
{}_2 F_1(-\nu,\nu+m;\frac{1}{2};x^2)
\label{sec3eq3}
\end{equation}
with the generalized Hypergeometric function, ${}_2 F_1$, the Bessel function, $J_{2n+1}$ and $y = -
(1-\lambda)\vec{\nabla}_1+\lambda\vec{\nabla}_2$. The series re-summation technique along with the 
{\it{Gegenbauer addition theorem}} \cite{watson} are used to arrive at the above expansion (i.e. 
Eq.(\ref{sec3eq1}) and Eq.(\ref{sec3eq2})). Now, Eq.(\ref{sec3eq1}) together with Eq.(\ref{sec2eq6}) 
produce the transformed density matrix
\begin{equation}
\varGamma_{1t}^{2D}=2\rho\frac{J_1(ku)}{ku}+\frac{6J_3(ku)}{k^3u}{\mathcal{G}}+\frac{24J_3(ku)}{k^3u^2}
{\mathcal{H}}~,
\label{sec3eq4}
\end{equation}
where
\begin{eqnarray}
{\mathcal{G}}&=&4\cos^2\phi\{(\lambda^2-\lambda+\frac{1}{2})\nabla^2\rho-2\tau\} + k^2\rho\nonumber\\
{\mathcal{H}}&=&\cos\phi(2\lambda-1)|\nabla\rho|
\label{sec3eq5}
\end{eqnarray}
with $\tau=\sum_{i}^{occ}|\vec{\nabla}\psi_{i}|^2$, the KS kinetic energy density. 
Now, to make $\tau$ gauge-invariant, we  modify it, so that
\begin{equation}
\tau\to \tilde{\tau}=\tau-2\frac{j_p^2}{\rho},
\label{exch9}
\end{equation}
where
\begin{equation}
j_p=\frac{1}{2i}\sum_{i}^{occ}\{\psi^*_i(\vec{r})[\vec{\nabla}\psi_i(\vec{r})]-[\vec{\nabla}\psi_i^*
(\vec{r})]\psi_i(\vec{r})\}
\label{exch10}
\end{equation}
is the para-magnetic current density. By doing this, the functional also becomes gauge invariant and 
fulfills all the above mentioned criteria. Inclusion of current density is particularly important whenever
there happens to be the radiation matter interactions. It is relevant to use the following cylindrical 
average of the exchange hole (e.g. shown in FIG.\ref{fig1}) corresponding to the above density matrix which will 
be used for the construction of the desired 2D semi-local functional i.e. 
\begin{equation}
\langle\rho^t_{x2D}\rangle=-\frac{2J_1^2(ku)}{k^2u^2}\rho(\vec{r})-\frac{24J_1(ku)J_3(ku)}{k^4u^2}
{\mathcal{L}}-\frac{144J_3^2(ku)}{k^6u^4}{\mathcal{M}}~,
\label{sec3eq6}
\end{equation} 
where
\begin{eqnarray}
{\mathcal{L}}&=&(\lambda^2-\lambda+\frac{1}{2})\nabla^2\rho-2\tau+4\frac{j_p^2}{\rho} + \frac{1}{2}k^2\rho
\nonumber\\{\mathcal{M}}&=&(2\lambda-1)^2\frac{|\nabla\rho|^2}{\rho}~.
\label{sec3eq7}
\end{eqnarray} 
By virtue of the above exchange hole, the functional retains the most unique features like uniform density 
limit for $k = k_F = (2\pi\rho)^{\frac{1}{2}}$ and correct  $u^2$ behavior. So the above DME based exchange 
hole is more general in nature than the previously proposed ones \cite{prvm}. In the earlier case \cite{prvm}, 
the small $u$ expansion of cylindrical average exchange hole was multiplied by the corresponding average 
exchange hole of uniform gas and the parameters were determined using the sum rule. Whereas, in the present 
attempt, all these are automatically taken care. Thus the uniform density limit is trivially recovered when 
$k = k_F$. But for inhomogeneous systems, the extent of inhomogeneity is included through a parameter $f$ 
(to be determined analytically) so that $k = fk_F$. Then, $f$ is being obtained from the normalization of 
the cylindrically averaged exchange hole i.e. 
\begin{equation}
\frac{1}{f^2}+\frac{6}{f^4}y = 1~,
\label{sec3eq8}
\end{equation} 
where $y=(2\lambda-1)^2p$ and $p=s^2=\frac{|\vec{\nabla}\rho|^2}{(2k_F\rho)^2}$ is the square of the reduced 
density gradient in 2D. For slowly varying density limit, Eq.(\ref{sec3eq8}) demands that $f \approx 1 +6 y$  
and in the limit of large density gradient, $f \to y^{\frac{1}{4}}$, similar to that proposed in \cite{prvm}. 
By applying successive root finding method to solve Eq.(\ref{sec3eq8}), we propose that for any arbitrary 
density the dimensionless parameter $f$ satisfies the relation
\begin{equation}
f = [1+90(2\lambda-1)^2p+\beta(2\lambda-1)^4p^2]^{\frac{1}{15}}~,
\label{sec3eq9}
\end{equation}
where $\beta$ is the parameter that has to be determined along with $\lambda$ by fitting with exact results 
known for physical systems. In this case, we found these parameters by comparing with exact exchange results
for the few electrons quantum dots. It is noteworthy to mention that in case for low density, the binomial 
expansion of Eq.(\ref{sec3eq9}) leads to $f \approx 1 +6 y$. Thus, the proposition, Eq.(\ref{sec3eq9}) 
is in right spirit. But the laplacian present in the exchange hole expansion also need to be removed in order 
to handle it numerically at the origin. The usual way to do so is to use the method of integration by parts. 
But here, we have used the semi-classical approximation of kinetic energy density \cite{bvz} to replace it. 
As this method has been successfully employed in designing the meta-GGA type functional in 3D \cite{tpss,tm16}. 
So $\nabla^2 \rho$ is replaced by
\begin{equation}
\nabla^2\rho=6\Big[\tau-\tau^{unif}_{2D}-2\frac{j_p^2}{\rho}\Big]~,
\label{sec3eq10}
\end{equation}
where $\tau^{unif}_{2D}=\frac{\pi\rho^2}{2}$. Thus, the modified exchange hole takes the form
\begin{equation}
\begin{split}
\langle\rho^t_{x2D}\rangle=-\frac{2J_1^2(fk_Fu)}{f^2k_F^2u^2}\rho(\vec{r})-\frac{24J_1(fk_Fu)J_3(fk_Fu)}
{f^4k_F^4u^2}{\mathcal{L}}\nonumber\\
-\frac{144J_3^2(fk_Fu)}{f^6k_F^6u^4}{\mathcal{M}}
\label{sec3eq11}
\end{split}
\end{equation} 
with
\begin{eqnarray}
{\mathcal{L}}&=&6(\lambda^2-\lambda+\frac{1}{2})\Big[\tau-\tau^{unif}_{2D}-2\frac{j_p^2}{\rho}\Big]
-2\tau +4\frac{j_p^2}{\rho}+ \frac{1}{2}k_F^2\rho\nonumber\\
{\mathcal{M}}&=&(2\lambda-1)^2\frac{|\nabla\rho|^2}{\rho}~.
\label{sec3eq12}
\end{eqnarray}
Now, from Eq.(\ref{sec2eq1}) and Eq.(\ref{sec3eq12}), the semilocal exchange energy density functional in
2D is given by
\begin{equation}
E^{2D-mGGA}_x=-\int\rho(\vec{r})\epsilon_x^{2D-LDA}F_x^{2D-mGGA}[p,\tau,j_p]~d^2r~,
\label{sec4eq1}
\end{equation}
where $\epsilon_x^{2D-LDA}=\frac{4k_F}{3\pi}$ and the enhancement factor (e.g. shown in FIG.\ref{fig2}),
\begin{equation}
F_x^{2D-mGGA}[p,\tau,j_p]=\frac{1}{f}+\frac{2R}{5f^3}
\label{sec4eq2}
\end{equation}
with
\begin{eqnarray}
R &=& 1 + \frac{128}{21}(2\lambda-1)^2p\nonumber\\
&+& \frac{3\Big(\lambda^2-\lambda + \frac{1}{2}\Big)\Big(\tau-\tau_{2D}^{unif}
-2\frac{j_p^2}{\rho}\Big)-\tau+2\frac{j_p^2}{\rho}}{\tau_{2D}^{unif}}~.
\label{sec4eq3}
\end{eqnarray}
\begin{table}[t]
\caption{Shown below are the exchange energies (in atomic units) for parabolically confined few electron 
quantum dots. The $1^{st}$ and $2^{nd}$ columns contain the number of particles and confinement strengths 
used for finding the parameters of the proposed functional. Results for EXX, 2D-LDA, 2D-GGA \cite{prvm}, 
2D-B88 \cite{vrmp} and 2D-BR \cite{prhg} are also shown for comparison with that obtained using the constructed 
2D-mGGA functional. The last row contains the mean percentage error, $\Delta$.}
\begin{tabular}{c  c  c  c  c  c   c   c}
\hline\hline
N&$\omega$&$-E_x^{EXX}$&$-E_{x}^{LDA}$&$-E_x^{GGA}$&$-E_x^{B88}$&$-E_x^{BR}$&$-E_x^{mGGA}$\\ \hline
2& 1/6&0.380&0.337& 0.368&0.364 &0.375 &0.386   \\
2& 0.25&0.485&0.431 & 0.470&0.464&0.480&0.492   \\
2& 0.50&0.729&0.649 & 0.707&0.699 &0.722&0.735   \\
2& 1.00& 1.083&0.967& 1.051 &1.039 &1.080&1.085 \\
2& 1.50&1.358&1.214& 1.319&1.304 &1.354 &1.354   \\
2& 2.50&1.797&1.610&1.748&1.728 &1.794 &1.776   \\
2& 3.50&2.157&1.934&2.097&2.074 &2.020 &2.113   \\
6& $1/1.89^2$&1.735&1.642&1.719&1.749&1.775&1.736  \\
6& 0.25&1.618&1.531 &1.603&1.594 &1.655 &1.620    \\
6& 0.42168&2.229&2.110 &2.206&2.241 &2.281 &2.226    \\
6& 0.50&2.470&2.339&2.444&2.431&2.529&2.466   \\
6& 1.00&3.732&3.537&3.690&3.742&3.824&3.716 \\
6& 1.50&4.726&4.482&4.672&4.648&4.845&4.699   \\
6& 2.50&6.331&6.008&6.258&6.226&6.492&6.279   \\
6& 3.50&7.651&7.264&7.562&7.525&7.846&7.573   \\
12&0.50&5.431&5.257&5.406&5.387&5.728&5.415    \\
12& 1.00&8.275&8.013& 8.230&8.311&8.572&8.231 \\
12& 1.50&10.535&10.206&10.476&10.444&10.915&10.461   \\
12& 2.50&14.204&13.765&14.122&14.080&14.716&14.063   \\
12& 3.50&17.237&16.709&17.136&17.086&17.858&17.019   \\
20&0.50&9.765&9.553&9.746&9.722&10.167&9.805    \\
20& 1.00&14.957&14.638&14.919& 15.029&15.573&14.894 \\
20& 1.50&19.108&18.704&19.053&19.188&19.892&19.007   \\
20& 2.50&25.875&25.334&25.796&25.973&26.935&25.698   \\
20& 3.50&31.491&30.837&31.392&31.603&32.777&31.230   \\
\hline
$\Delta$& & &5.7&1.7 &3.9 &2.8&{\bf{0.7}}\\
\hline\hline
\end{tabular}
\end{table}
To test the functional and obtain the parameters $\beta$ and $\lambda$, we have chosen the set of parabolic 
quantum dots having varying confinement strengths with few electrons embedded into it. This type of system 
is reported for testing the 2D-GGA functional \cite{prvm} and to fix the corresponding parameters involved
therein. A self-consistent calculation with KLI-OEP exact-exchange method using OCTOPUS code \cite{octopus} 
has been performed and the density is being used as the reference input. As our system is non-magnetic, so
$j_p=0$. The value of $\lambda$ is obtained by fitting with different confinement strengths such that the 
mean percentage error gets reduced. Whereas, $\beta$ is fixed so as to confirm the smooth behavior of the 
enhancement factor in the $s\approx 0$ region \cite{ptss04}. The value of the parameters $\lambda$ and 
$\beta$ are obtained to be $0.74$ and $30.0$ respectively. In $3D$ \cite{tm16}, same set of parameters are 
also used. But those are fixed by taking the exact exchange of hydrogen atom along with the smooth behavior 
of the enhancement factor at the iso-orbital region in order to remove the spurious divergence of the exchange 
potential \cite{ptss04}. The current functional is tested and the performance of it is shown in Table-I. 
Trivially, the results are quite superior as it yields error that are smaller by at least a factor of 
$8.1$, $2.4$, $5.6$ and $4.0$ w.r.t. 2D-LDA, 2D-GGA \cite{prvm}, 2D-B88 \cite{vrmp} and 2D-BR \cite{prhg} 
respectively. Lastly, the comprehensive assessment of the functional is being performed for Gaussian quantum 
dots by simultaneously varying the the number of electrons trapped $N$, depth of the potential and  confinement 
strength $\omega$. For this case, the performance is presented in Table-II and FIG.\ref{fig3}. Here too, the 
results are found to be in excellent agreement with KLI-EXX. Actually, the new semilocal functional reduces 
the error by a factor of $2.2$ compared to 2D-GGA for the whole set.
\begin{table}
\caption{Comparison of exchange energies (a.u.) for Gaussian quantum dots ($v_{ext}=-V_0e^{-\omega^2r^2})$ 
are shown for low density. Mean percentage error given in the last row.} 
\begin{tabular}{c  c  c  c  c  c  c  c  c }
\hline\hline
$V_0$ & N &$\omega^2$ & $-E_x^{EXX}$ & $-E_{x}^{LDA}$ & $-E_x^{GGA}$ & $-E_x^{mGGA}$\\ \hline
10    & 2 &0.05       &1.047         &0.934           &1.017         &1.048\\
10    & 2 &0.10       &1.255         &1.120           &1.219         &1.250\\
10    & 2 &0.25       &1.573         &1.405           &1.529         &1.555\\
10    & 2 &1/6        &1.427         &1.274           &1.386         &1.416\\
10    & 2 &0.50       &1.839         &1.643           &1.788         &1.804\\
40    & 6 &0.05       &5.416         &5.139           &5.354         &5.372\\
40    & 6 &0.10       &6.525         &6.194           &6.450         &6.460\\
40    & 6 &0.25       &8.255         &7.840           &8.160         &8.142\\
40    & 6 &1/6        &7.454         &7.076           &7.367         &7.364\\
\hline
$\Delta$& & & &8.3&2.0 &{\bf{0.9}}\\
\hline\hline
\end{tabular}
\end{table}

To summarize, a meta-GGA type semi-local functional in two dimensions is constructed based on DME. 
The beauty of this functional is that, the exchange hole involved in it has correct short range 
behavior and recovers the uniform density limit quite accurately. The convergence of the exchange hole in 
large separation limit leads to an analytical expression for the corresponding energy functional 
even without applying any cutoff procedure which are essentially lacked by 2D functionals 
proposed so far. The most appealing feature of the present semi-local functional is that it is 
derived from the full exchange hole and thus having strong physical basis. The functional is one 
step ahead of the 2D-GGA as it leads to significant reduction in error compare to it's counterparts. 
Thus, the functional in principle can enable us for making precise many-electron calculations of 
larger structures such as arrays of quantum-dots, quantum-Hall devices, semiconductor quantum dots, 
quantum Hall bars on a regular basis. Also, the constructed exchange hole can be used to construct 
meta-GGA level exchange only pair-distribution function, static structure factor, non-local and 
range separated functionals in 2D. The present construction can be further extended to the recently 
developed density functional formalism for strictly correlated electrons. The next step is to construct 
functional for correlation energy which will be compatible with the exchange. The functional is not 
only physically appealing but also practically useful as it opens the path for constructing exchange 
correlation functionals in two dimensions analogue to the Jacob's ladder in three dimensions.

The authors would like to acknowledge the financial support from the Department of Atomic Energy, 
Government of India.


\begin{thebibliography}{unsrt}
\bibitem{hk64} P. Hohenberg and W. Kohn, Phys. Rev. {\bf{136}}, B864 (1964).
\bibitem{ks65} W. Kohn and L. J. Sham, Phys. Rev. {\bf 140}, A1133 (1965).
\bibitem{b83} A. D. Becke, Int. J. Quant. Chem. {\bf 23}, 1915 (1983).
\bibitem{jp85} J. P. Perdew, Phys. Rev. Lett. {\bf{55}}, 1665 (1985)
\bibitem{pw86} J. P. Perdew and Y. Wang, Phys. Rev. B {\bf 33}, 8800 (1986).
\bibitem{b88} A. D. Becke, Phys. Rev. A {\bf 38}, 3098 (1988).
\bibitem{br89} A. D. Becke and M. R. Roussel, Phys. Rev. A {\bf 39}, 3761 (1989).
\bibitem{b3pw91} A. D. Becke, J. Chem. Phys. {\bf 104}, 1040 (1996).
\bibitem{pbe96} J. P. Perdew, K. Burke and M. Ernzerhof, Phys. Rev. Lett. {\bf 77}, 3865 (1996).
\bibitem{kos} R. M. Koehl, G. K. Odom and G. E. Scuseria, Mol. Phys. {\bf 87}, 835 (1996).
\bibitem{vsxc98}  T. V. Voorhis and G. E. Scuseria, J. Chem. Phys. {\bf 109}, 400 (1998).
\bibitem{hcth} F. A. Hamprecht, A. J. Cohen, D. J. Tozer and N. C. Handy, J. Chem. Phys. {\bf 109}, 
6264 (1998).
\bibitem{tsuneda} T. Tsuneda and K. Hirao, Phys. Rev. B {\bf 62}, 15527 (2000).
\bibitem{tpss} J. Tao, J. P. Perdew, V. N. Staroverov and G. E. Scuseria, Phys. Rev. Lett. {\bf 91}, 
146401 (2003).
\bibitem{mO6l} Y. Zhao and D. G. Truhlar, J. Chem. Phys. {\bf 125}, 194101 (2006).
\bibitem{revtpss} J. P. Perdew, A. Ruzsinszky, G.I. Csonka, L. A. Constantin and J. Sun, Phys. Rev. 
Lett. {\bf 103}, 026403 (2009). 
\bibitem{tbmbj} F. Tran and P. Blaha, Phys. Rev. Lett. {\bf{102}}, 226401 (2009).
\bibitem{scan15}  J. Sun, A. Ruzsinszky and J.P. Perdew, Phys. Rev. Lett. {\bf 115}, 036402 (2015).
\bibitem{tm16} J. Tao and Y. Mo, Phys. Rev. Lett. {\bf{117}}, 073001 (2016). 
\bibitem{kat} L. P. Kouwenhoven, D. G. Austing and S. Tarucha, Rep. Prog. Phys. {\bf{64}}, 701 (2001).
\bibitem{rm} S. M. Reimann and M. Manninen, Rev. Mod. Phys. {\bf{74}}, 1283 (2002).
\bibitem{klnlhm} Y. -H. Kim, I. -H. Lee, S. Nagaraja, J.-P. Leburton, R. Q. Hood and R. M. Martin, 
Phys. Rev. B {\bf{61}}, 5202 (2000).
\bibitem{ccfs} L. Chiodo, L. A. Constantin, E. Fabiano and F. Della Sala, Phys. Rev. Lett. {\bf 108}, 
126402 (2012).
\bibitem{rk} A. K. Rajagopal and J. C. Kimball, Phys. Rev. B {\bf{15}}, 2819 (1977).
\bibitem{tc} B. Tanatar and D. M. Ceperley, Phys. Rev. B {\bf{39}}, 5005 (1989).
\bibitem{amgb} C. Attaccalite, S. Moroni, P. Gori-Giorgi and G. B. Bachelet, Phys. Rev. Lett. 
{\bf{88}}, 256601  (2002).
\bibitem{hser} H. Saarikoski, E. R\"{a}s\"{a}nen, S. Siljamäki, A. Harju, M. J. Puska and R. M. Nieminen,
Phys. Rev. B {\bf{67}}, 205327 (2003).
\bibitem{prhg} S. Pittalis, E. R\"{a}s\"{a}nen, N. Helbig and E. K. U. Gross, Phys. Rev. B 
{\bf{76}}, 235314 (2007).
\bibitem{prvm} S. Pittalis, E. R\"{a}s\"{a}nen, J. G. Vilhena and M. A. L. Marques,
Phys. Rev. A {\bf{79}}, 012503 (2009).
\bibitem{prg} S. Pittalis, E. R\"{a}s\"{a}nen and E. K. U. Gross, Phys. Rev. A {\bf{80}}, 032515 (2009).
\bibitem{pr1} S. Pittalis and E. R\"{a}s\"{a}nen, Phys. Rev. B {\bf{80}}, 165112 (2009).
\bibitem{prp} S. Pittalis, E. R\"{a}s\"{a}nen and C. R. Proetto, Phys. Rev. B {\bf{81}}, 115108 (2010).
\bibitem{rp} E. R\"{a}s\"{a}nen and S. Pittalis, Physica E {\bf{42}}, 1232–1235 (2010).
\bibitem{sr} S. Sakiroglu and E. R\"{a}s\"{a}nen, Phys. Rev. A {\bf{82}}, 012505 (2010).
\bibitem{pr2} S. Pittalis and E. R\"{a}s\"{a}nen, Phys. Rev. B {\bf{82}}, 165123 (2010).
\bibitem{rpvm} E. R\"{a}s\"{a}nen, S. Pittalis, J. G. Vilhena and M. A. L. Marques, 
Int. J. Quant. Chem. {\bf{110}}, 2308–2314 (2010).
\bibitem{prlvm} A. Putaja, E. R\"{a}s\"{a}nen, R. van Leeuwen, J. G. Vilhena and M. A. L. Marques,
Phys. Rev. B {\bf{85}}, 165101 (2012).
\bibitem{vrmp} J. G. Vilhena,E. R\"{a}s\"{a}nen, M. A. L. Marques and S. Pittalis,
J. Chem. Th. Comp. {\bf{10}}, 1837−1842 (2014).
\bibitem{prm} S. Pittalis, E. R\"{a}s\"{a}nen and M. A. L. Marques, Phys. Rev. B {\bf{78}}, 195322 (2008). 
\bibitem{prpg} S. Pittalis, E. R\"{a}s\"{a}nen, C. R. Proetto and E. K. U. Gross, Phys. Rev. B {\bf{79}}, 
085316 (2009).
\bibitem{rpp} E. R\"{a}s\"{a}nen, S. Pittalis and C. R. Proetto, Phys. Rev. B {\bf{81}}, 195103 (2010).
\bibitem{nv1} J. W. Negele and D. Vautherin, Phys. Rev. C {\bf 5}, 1472 (1972).
\bibitem{kli} J. B. Krieger, Y. Li, and G. J. Iafrate, Phys. Rev. A {\bf{46}}, 5453 (1992).
\bibitem{hkprg} N. Helbig, S. Kurth, S. Pittalis, E. R\"{a}$s$\"{a}nen and E. K. U. Gross, 
Phys. Rev. B {\bf{77}}, 245106 (2008).
\bibitem{tsp03} J. Tao, M. Springborg and J. P. Perdew, J. Chem. Phys. {\bf 119}, 6457 (2003).
\bibitem{watson} G. N. Watson, A Treatise on the Theory of Bessel Functions, New York: Macmillan, (1944).
\bibitem{bvz} M. Brack and B. P. van Zyl, Phys. Rev. Lett. {\bf{86}}, 1574 (2001).
\bibitem{octopus} M. A. L. Marques, A. Castro, G. F. Bertsch and A. Rubio, Comp. Phys. Comm. 
{\bf{151}}, 60 (2003).
\bibitem{ptss04} J. P. Perdew, J. Tao, V. N. Staroverov and G.E. Scuseria, J. Chem. Phys.  
{\bf 120}, 6898 (2004).
\end{thebibliography}
\end{document}